\documentclass[11pt]{article}  
\usepackage{times,amsmath,amsfonts,amsthm,amssymb,eucal}
\usepackage{array}
\usepackage{color}
\usepackage{enumerate}
\usepackage{epsfig}
\usepackage{mathrsfs}
\usepackage{fancyhdr}
\usepackage{graphicx}
\usepackage{caption}
\usepackage{subcaption}
\usepackage{graphicx}
\usepackage{makeidx}
\usepackage{multirow}
\theoremstyle{plain}
\newtheorem{theorem}{Theorem}
\newtheorem{proposition}{Proposition}
\newtheorem{lemma}{Lemma}
\newtheorem{corollary}{Corollary}
\newtheorem{definition}{Definition}

\newtheorem{remark}{Remark}

\usepackage{secdot}
\sectiondot{subsection}

\usepackage[center,md]{titlesec}

\theoremstyle{definition}

\newcommand{\E}{\mathrm{E}}
\def\Theorem{\begin{theorem}\sl}
\def\EndTheorem{\end{theorem}}
\def\Proposition{\begin{proposition}\sl}
\def\EndProposition{\end{proposition}}
\def\Lemma{\begin{lemma}\sl}
\def\EndLemma{\end{lemma}}
\def\Remark{\begin{remark}\sl}
\def\EndRemark{\end{remark}}
\def\Corollary{\begin{corollary}\sl}
\def\EndCorollary{\end{corollary}}
\def\Definition{\begin{definition}\sl}
\def\EndDefinition{\end{definition}}
\numberwithin{equation}{section}

\begin{document}
\title{ \textbf{ESTIMATION OF POVERTY MEASURES FOR SMALL AREAS UNDER A TWO-FOLD NESTED ERROR LINEAR REGRESSION MODEL: COMPARISON OF TWO METHODS}}
\author{Maryam Sohrabi, J.N.K. Rao}
\date{Statistics Canada and Carleton University}
\maketitle

\noindent{\large\emph{ABSTRACT:}} Demand for reliable statistics at a local area (small area) level has greatly increased in recent years. Traditional area-specific estimators based on probability samples are not adequate because of small sample size or even zero sample size in a local area. As a result, methods based on models linking the areas are widely used. World Bank focused on estimating poverty measures, in particular poverty incidence and poverty gap called FGT measures, using a simulated census method, called ELL, based on a one-fold nested error model for a suitable transformation of the welfare variable. Modified ELL methods leading to significant gain in efficiency over ELL also have been proposed under the one-fold model. An advantage of ELL and modified ELL methods is that distributional assumptions on the random effects in the model are not needed. In this paper, we extend ELL and modified ELL to two-fold nested error models to estimate poverty indicators for areas (say a state) and subareas (say counties within a state). Our simulation results indicate that the modified ELL estimators lead to large efficiency gains over ELL at the area level and subarea level. Further, modified ELL method retaining both area and subarea estimated effects in the model (called MELL2) performs significantly better in terms of mean squared error (MSE) for sampled subareas than the modified ELL retaining only estimated area effect in the model (called MELL1).\\ 

\noindent\emph{Keywords and phrases:} Areas and subareas, ELL and modified ELL methods, Poverty incidence and gap, Two-fold nested error model.

\vspace{5mm}

\begin{center}
     \section{INTRODUCTION}
\end{center}

Data collected from probability samples can provide reliable estimates of parameters of interest for domains (subpopulations) with large enough sample sizes to permit direct, domain-specific estimators of desired precision. We call such domains as large areas. On the other hand, sample sizes can be very small or even zero for local areas (called small areas) and direct estimators are not adequate or feasible. Demand for reliable statistics at the level of small areas has increased greatly and it is necessary to use model-based methods that can yield reliable estimates for small areas by integrating information across areas through linking models. Rao and Molina (2015) provide a comprehensive account of model-based small area estimation of means, totals, and more complex parameters like poverty measures.

In this paper, we focus on the estimation of FGT poverty measures, proposed by Foster, Greer and Thorbeck (1984). Poverty incidence, gap and severity belong to the family of FGT measures. World Bank widely used a method proposed by Elbers, Lanjouw and Lanjouw (2003), called the ELL method, to provide FGT poverty measures for specified local areas in many developing countries. The ELL method involves the following steps: (1) Simulate multiple censuses of the welfare variable of interest based on an assumed model relating the welfare variable to auxiliary variables obtained from a recent census. (2) Calculate the FGT measure for specified local areas from each simulated census and then take the average over the censuses as the ELL estimator. (3) Variance of the simulated census estimators is taken as the estimator of mean squared error (MSE) of the ELL estimator. An advantage of the ELL method is that it is free of parametric distributional assumptions and computationally simple. However, Molina and Rao (2010) showed that the ELL method can lead to large MSE compared to an optimal method, called the Empirical Best (EB) method, assuming a one-fold nested error linear regression model with normally distributed random effects. Diallo and Rao (2018) developed a modification to ELL method that leads to substantial reduction in MSE and compares favorably to the normality-based EB method. As in the ELL method, the modified ELL method is free of parametric distributional assumptions.                                                                                                     
The proposed ELL, modified ELL and EB methods are based on a one-fold nested error linear regression model relating a suitable function of the welfare variable to the census variables and a random area effect. Sample survey data observing the welfare variable and the census variables, based on two-stage cluster sampling, are used to fit the one-fold model. In the traditional ELL method, random cluster effects are included in the model and simulated censuses are generated. From a simulated census, a desired poverty measure is calculated for any desired small area. Note that it is not necessary to specify the areas in advance because area effects are not included in the ELL one-fold model. Hossain et al. (2020) used a two-stage sample of districts and household within districts to estimate a food insecurity measure at the district level in Bangladesh. In this case, clusters are areas. 
In this paper, we focus on two-fold random effect models involving area and subarea random effects. For example, an area could refer to a state and a subarea to a county within a state. Marhuenda et al. (2017) studied EB estimation of FGT poverty measures under the two-fold model, assuming that the random effects in the model are normally distributed, as in the case of the one-fold model studied by Molina and Rao (2010). In their application to Spanish survey data, areas are provinces and subareas are comarcas, and it is of interest to obtain estimates of poverty measures at the domain as well as subdomain level. Section 2 introduces the  two-fold model and the associated FGT poverty measures for domains and subdomains. Section 3 extends the ELL and modified ELL methods to two-fold models with no distributional assumptions on the random effects in the model, as in the case of the one-fold model. Section 4 presents some results of a simulation study on the performance of ELL and modified ELL estimators. Finally, some remarks on the estimation of MSE of the estimators are given in Section 5. 

\vspace{5mm}

\begin{center}
\section{ TWO-FOLD NESTED ERROR MODEL } 
\end{center}

The finite population of interest consists of $D$ areas (domains)  $d=1,\ldots,D$, and area $d$ is divided into $M_d$ subareas (subdomains)  $j=1,\ldots,M_d$.  The subdomain $j$  within the domain  $d$ contains $N_{dj}$ elements $k=1,\ldots,N_{dj}$. The population data is denoted as  $\{(E_{djk},\mathbf{x}_{djk}^T),d=1,\ldots,D;j=1,\ldots,M_d;k=1,\ldots,N_{dj}\}$, where $E_{djk}$ is the welfare variable of interest and  $\mathbf{x}_{djk}^{T}=(x_{1djk},\ldots,x_{pdjk})$   is a  $p$-vector of known census variables. If an intercept term is needed, then we set ${x}_{1djk}=1$  for all the population units. To reduce positive skewness of the welfare variable we make a log transformation  $y_{djk}=\log(E_{djk})$. 

A two-fold nested error population model relating the transformed variable $y_{djk}$  to the census variables $\mathbf{x}_{djk}$ is given by 

\begin{align}
\label{TF}
y_{djk}=\mathbf{x}_{djk}^T\boldsymbol{\beta}+u_d+v_{dj}+e_{djk}; \ \ \  d=1,\ldots,D, \ j=1,\ldots,M_d, \ k=1,\ldots,N_{dj},
\end{align}
where  $\boldsymbol{\beta}$ is a $p\times 1$ vector of unknown regression parameters, $u_d$ are the area effects, $v_{dj}$ are the cluster effects, and $e_{djk}$ are the residual errors. The three random errors $u_d$, $v_{dj}$, and $e_{djk}$ are independent with $\E(u_d)=\E(v_{dj})=\E(e_{djk}) = 0$. Parametric distributions on the two random effects and the unit errors are not assumed.

We assume two-stage sampling in each area:  a sample, $s_d$, of $m_d (\leq M_d)$ subareas is selected from area  $d$ and if subarea  $j$th  is sampled then a subsample, $s_{dj}$ of $n_{dj}$ elements is selected from subarea $j$. We further assume that the population model \eqref{TF} also holds for the sample data  $\{(y_{djk},\mathbf{x}_{djk}),d=1,\ldots,D;j=1,\ldots,m_d;k=1,\ldots,n_{dj}\}$. Therefore, the model for sample data is given by 
\begin{align}
\label{TFS}
y_{djk}=\mathbf{x}_{djk}^T\boldsymbol{\beta}+u_d+v_{dj}+e_{djk}; \ \ \  d=1,\ldots,D, \ j=1,\ldots,m_d, \ k=1,\ldots,n_{dj}.
\end{align}
The FGT population measure for area $d$ is given by 
\begin{align}
\label{FGT}
F_{\alpha d}(z)=\frac{1}{N_d}\sum_{j=1}^{M_d}\sum_{k=1}^{N_{dj}}F_{\alpha djk},
\end{align}
where  $N_d=\sum_{j}N_{dj}$ and
\begin{align}
\label{F}
F_{\alpha djk}=\left(\frac{z-E_{djk}}{z}\right)^{\alpha}I(E_{djk}<z).
\end{align}
In \eqref{F},  $z$ is the known poverty line and $I(E_{djk}<z)$  is the indicator variable taking the value $1$   when  $E_{djk}$  is smaller than  $z$  and  0  otherwise. Poverty incidence, poverty gap and poverty severity correspond to $\alpha=0$, $\alpha=1$,  and  $\alpha=2$, respectively. Also, the FGT measure for subarea $j$ within area $d$ is given by
\begin{align}
\label{FGTS}
F_{\alpha dj}(z)=\frac{1}{N_{dj}}\sum_{k=1}^{N_{dj}}F_{\alpha djk}.
\end{align}
 
\vspace{5mm}

\begin{center}
\section{ESTIMATORS OF FGT POVERTY MEASURES}
\end{center}

In this section, we describe how to estimate FGT poverty measures \eqref{FGT} and \eqref{FGTS} for areas and subareas, respectively. Suppose that there is a one-to-one transformation $y_{djk}=\log(E_{djk})$ of the welfare variables, $E_{djk}$. Then we can express  $F_{\alpha djk}$ given as \eqref{F} in terms of $y_{djk}$:
\begin{align*}
F_{\alpha djk}=\left(\frac{z-\exp(y_{djk})+c}{z}\right)^{\alpha}I(\exp(y_{djk})-c<z):=h_{\alpha}(y_{djk}).
\end{align*}

\vspace{5mm}

\noindent\textbf{3.1. \quad{ELL Method}}\\
\label{ELLO}
\vspace{2mm}

Elbers, Lanjow and Lanjow (2003) consider a  linear mixed model for a log-transformation of  the variable measuring welfare of individuals, with random effects for the sampling clusters. In small area context,  we can assume that the sampling clusters are the
areas. In this case, the model becomes the one-fold nested error model of Battese, Harter
and Fuller (1988)  for the log-transformation of the welfare variables, that is, $y_{dj }=\log(E_{dj})$.  The World Bank applied the ELL  method  extensively to obtain   poverty and inequality measures for many countries: for more details see Elbers, Lanjow and Lanjow (2003). 
In this section, we extend the ELL method under the one-fold nested error model to the two-fold nested error model \eqref{TF}  when the model has area level random effect term,  subarea level random effect term, and  subarea level error term. 
The ELL method consists of drawing from the estimated area, cluster  and unit level residuals to
create a simulated census. The steps of the ELL method can be summarized as follows:\\

  \begin{enumerate}
  \item Estimate $\boldsymbol{\beta}$ from the nested error model given by \eqref{TFS}, using the  ordinary least squares (OLS) method, and obtain unit level residuals $\hat{r}_{djk}=y_{djk}-\mathbf{x}_{djk}^T\hat{\boldsymbol{\beta}}_{OLS}$, where $\hat{\boldsymbol{\beta}}_{OLS}$ denotes the estimator of $\boldsymbol{\beta}$."
  \item The area effect ${u}_{d}$, the subarea effect ${v}_{dj}$, and the unit level errors ${e}_{djk}$ are estimated as $$\hat{u}_{d}=\frac{1}{n_d}\sum_{j=1}^{m_d}\sum_{k=1}^{n_{dj}}\hat{r}_{djk},$$
  $$\hat{v}_{dj}=\frac{1}{n_{dj}}\sum_{k=1}^{n_{dj}}\hat{r}_{djk}-\hat{u}_{d},$$
  and
  $$\hat{e}_{djk}=\hat{r}_{djk}-\frac{1}{n_{dj}}\sum_{k=1}^{n_{dj}}\hat{r}_{djk}.$$
  \item Draw  $\hat{\boldsymbol{\beta}}^{*(b)}$, ${u}_d^{*(b)}$, ${v}_{dj}^{*(b)}$, and ${e}_{djk}^{*(b)}$, $b=1,\ldots,B$ from $N(\hat{\boldsymbol{\beta}}_{OLS}, Cov(\hat{\boldsymbol{\beta}}_{OLS}))$, the empirical distribution of $\hat{u}_d$, the empirical distribution of $\hat{v}_{dj}$, and the empirical distribution of $\hat{e}_{djk}$, respectively.
  \item Construct $B$ simulated census values $\{y_{ djk}^{*(b)}; k=1,\ldots,N_{dj},j=1,\ldots,M_d,d=1,\ldots,D\}$  as follows:
  ${y}_{ djk}^{*(b)}=\mathbf{x}_{djk}^T\hat{\boldsymbol{\beta}}^{*(b)}+{u}_d^{*(b)}+{v}_{dj}^{*(b)}+{e}_{djk}^{*(b)}$, using the census values of the covariates.  
 \item The simulated population measures ${F}_{\alpha d}^{*(b)}=\frac{1}{N_d}\sum_{j=1}^{M_d}\sum_{k=1}^{N_{dj}}F_{\alpha djk}^{*(b)}$ and  ${F}_{\alpha dj}^{*(b)}=\frac{1}{N_{dj}}\sum_{k=1}^{N_{dj}}F_{\alpha djk}^{*(b)}$ are calculated from each simulated census $b$, where 
${F}^{*(b)}_{\alpha djk}= h_\alpha(y_{ djk}^{*(b)})$, $b = 1, \ldots , B$. 
\item The ELL estimators of $F_{\alpha d}$ and $F_{\alpha dj}$ are calculated   by
averaging over the $B$  simulated measures as follows:
  \begin{align*}
  \hat{F}_{\alpha d}^{ELL}=\frac{1}{B}\sum_{b=1}^{B} {F}_{\alpha d}^{*(b)}
  \end{align*}
  and
    \begin{align*}
  \hat{F}_{\alpha dj}^{ELL}=\frac{1}{B}\sum_{b=1}^{B} {F}_{\alpha dj}^{*(b)}.
  \end{align*}
\end{enumerate}
 
 \vspace{5mm}
\label{ELLM}
\noindent\textbf{3.2 \quad Modified ELL}\\

\vspace{2mm}

\noindent{\bf Method 1.}  This modification retains $\hat{u}_d$  in constructing the predictors $y_{djk}^{*(b)}$, unlike the use of ${u}_d^{*(b)}$   in the ELL method. We have the following modified ELL method:

\begin{enumerate}
  \item From the nested error model given by \eqref{TFS}, estimate the fixed effects $\boldsymbol{\beta}$ using OLS.
\item Estimate ${u}_{d}$, ${v}_{dj}$, and ${e}_{djk}$ as in the traditional ELL method.
\item Draw   ${v}_{dj}^{*(b)}$ and ${e}_{djk}^{*(b)}$, $b=1,\ldots,B$ from the empirical distributions  of  $\hat{v}_{dj}$ and  $\hat{e}_{djk}$, respectively.
  \item Construct $B$ simulated census values $\{y_{ djk}^{*(b)}; k=1,\ldots,N_{dj},j=1,\ldots,M_d,d=1,\ldots,D\}$ as follows:
  $${y}_{ djk}^{*(b)}=\mathbf{x}_{djk}^T\hat{\boldsymbol{\beta}}_{OLS}+\hat{u}_d+{v}_{dj}^{*(b)}+{e}_{djk}^{*(b)}.$$
   \item  Then, the simulated population measures ${F}_{\alpha d}^{*(b)}$ and  ${F}_{\alpha dj}^{*(b)}$ are calculated  as in the traditional ELL method from
each simulated census $b$.  The modified ELL estimators of $F_{\alpha d}$ and  $F_{\alpha dj}$, denoted by $\hat{F}_{\alpha d}^{MELL1}$ and $\hat{F}_{\alpha dj}^{MELL1}$, respectively, are as follows:
  \begin{align*}
  \hat{F}_{\alpha d}^{MELL1}=\frac{1}{B}\sum_{b=1}^{B} {F}_{\alpha d}^{*(b)}
  \end{align*}
  and 
   \begin{align*}
  \hat{F}_{\alpha dj}^{MELL1}=\frac{1}{B}\sum_{b=1}^{B} {F}_{\alpha dj}^{*(b)}.
  \end{align*}
 \end{enumerate}

\noindent{\bf Method 2.}  This modification retains $\hat{u}_d$ and $\hat{v}_{dj}$, for $j \in s_d$, and uses  ${v}_{dj}^{*(b)}$ for  subarea $j$ not sampled from area $d$ in constructing the predictors $y_{djk}^{*(b)}$. Then, the modification is as follows:

\begin{enumerate}
  \item From the nested error model given by \eqref{TFS}, estimate the fixed effects $\boldsymbol{\beta}$ using OLS.
\item Estimate ${u}_{d}$, ${v}_{dj}$, and ${e}_{djk}$ as in the traditional ELL method.
\item Draw  ${e}_{djk}^{*(b)}$, $b=1,\ldots,B$ from the empirical distribution of $\hat{e}_{djk}$.
  \item Construct $B$ simulated census values $y_{ djk}^{*(b)}$ for the units in the sampled subareas as
  $${y}_{ djk}^{*(b)}=\mathbf{x}_{djk}^T\hat{\boldsymbol{\beta}}_{OLS}+\hat{u}_d+\hat{v}_{dj}+{e}_{djk}^{*(b)}$$
  and for subareas that are not sampled ${y}_{ djk}^{*(b)}$  are generated  from
  $${y}_{ djk}^{*(b)}=\mathbf{x}_{djk}^T\hat{\boldsymbol{\beta}}_{OLS}+\hat{u}_d+{v}_{dj}^{*(b)}+{e}_{djk}^{*(b)},$$
   where  ${v}_{dj}^{*(b)}$,  $b=1,\ldots,B$, are drawn from the empirical distribution  $\hat{v}_{dj}$.
    \item  Then, the simulated population measures ${F}_{\alpha d}^{*(b)}$ and ${F}_{\alpha dj}^{*(b)}$ are calculated as in the traditional ELL method from
each simulated census $b$, and the second modified ELL estimators of $F_{\alpha d}$ and $F_{\alpha dj}$, denoted by $\hat{F}_{\alpha d}^{MELL2}$ and $\hat{F}_{\alpha dj}^{MELL2}$, respectively, are as follows:
  \begin{align*}
  \hat{F}_{\alpha d}^{MELL2}=\frac{1}{B}\sum_{b=1}^{B} {F}_{\alpha d}^{*(b)}
  \end{align*}
  and 
   \begin{align*}
  \hat{F}_{\alpha dj}^{MELL2}=\frac{1}{B}\sum_{b=1}^{B} {F}_{\alpha dj}^{*(b)}.
  \end{align*}
\end{enumerate}

\vspace{5mm}

\begin{center}
\section{SIMULATION STUDY}
\end{center}

A simulation study is undertaken to examine the performance of the two modified ELL methods under the two-fold nested error linear regression model \eqref{TF}. Marhuenda et al. (2017) conducted a simulation study on the performance of EB estimators of FGT measures for areas and subareas under a two-fold nested error model assuming  $u_d$, $v_{dj}$ and $e_{djk}$ are normally distributed. We follow their simulation set-up but also consider skew normal scenarios: (1)  $(u_d,v_{dj})$ normal (N) and  $e_{djk}$ skew normal (SN). (2)  $u_d$ normal and $(v_{dj},e_{djk})$ skew normal. Section 4.1 reports results for case 1 and results for case 2 are given in section 4.2. We also include the case of  $(u_d,v_{dj},e_{djk})$  normal (N) studied by Marheunda et al. (2017).

We generated  $I = 1000$ populations each of size $N = 20,000$  composed of  $D = 40$ areas each containing  $M_d = 10$  subareas each containing $N_{dj} = 50$  units. We first generated the covariate vector  $\mathbf{x}_{djk}=(1,x_{1djk},x_{2djk})^{'}$   for each population unit, based on $x_{1djk} \sim\mbox{B}(1,p_{1dj})$  and $x_{2djk} \sim\mbox{B}(1,p_{2dj})$ with probabilities  $p_{1dj}=0.2+\frac{0.4d}{D}+\frac{0.4j}{M_d}$ and  $p_{2dj}=0.2, \ j = 1, \ldots, 10, \ d = 1,\ldots , 40$. The generated population covariate values are held fixed and used to generate the dependent variable  $y_{djk}$ from the two-fold model using  $\beta = (3, 0.03,-0.04)^{'}$ and specified distributions for  $u_d$, $v_{dj}$, and $e_{djk}$   with mean zero and standard deviations   $\sigma_u=0.5$,  $\sigma_v=0.25$ and $\sigma_e=0.50$, respectively. In case 1, $u_d\sim\mbox{N}(0,\sigma_u^2)$, $v_{dj}\sim\mbox{N}(0,\sigma_v^2)$ and $e_{djk}\sim\mbox{SN}(\mu,\sigma^2,\lambda)$  with $\mu$  and  $\sigma$  chosen to make the mean and standard deviation of $e_{djk}$ equal to zero and  $\sigma_e$, and  $\lambda=\lambda_e=3$ which leads to moderate skewness. Note that the two- fold model is applied to $y_{djk}=\log(E_{djk})$  which reduces the skewness in the welfare variable  $E_{djk}$. As a result, moderate skewness in the errors $e_{djk}$  is realistic. In case 2,  $u_d\sim\mbox{N}(0,\sigma_u^2)$ and $(v_{dj},e_{djk})$ skew normal with mean zero and standard deviations   $\sigma_v=0.25$ and $\sigma_e=0.50$   respectively, and  $\lambda_v=1$ and $\lambda_e=3$, respectively.  The above process was repeated to generate   $I = 1000$ population values $\{y_{djk},i = 1,\ldots , 1000\}$.  

We calculated the FGT measures  ${F}_{\alpha d}^{(i)}$ for each area and  ${F}_{\alpha dj}^{(i)}$ for each subarea from each of the simulated populations $i = 1,\ldots , 1000$.  We focus on poverty incidence $(\alpha=0)$  and poverty gap $(\alpha=1)$ . Following Marheunda et al. (2017), we took the poverty line as  $z=0.6\mbox{med}(E_{djk})$ for a population generated as above, where $E_{djk}=\exp(y_{djk})$.

We considered two cases for generating a sample of units. In case I, all subareas are sampled $(m_d=M_d=10)$  by selecting  $n_{dj}=10$ units from each subarea by simple random sampling. In case II, a simple random sample of  $m_d=5$ subareas is selected from each area and then a simple random sample of $n_{dj}=20$ units is drawn from each sampled subarea. In both cases, the over all sample size within each area is equal to  $100$. 

We used a model-based set up by conditioning on the selected sample of units and extracting the corresponding sample data $(y_{djk}^{(i)},\mathbf{x}_{djk})$  from each simulated population  $i$. Using the sample data, we then obtained the desired estimates for areas and subareas from the assumed two-fold model. Denoting the estimators for areas and subareas for any given method by $\hat{F}_{\alpha d}$  and  $\hat{F}_{\alpha dj}$ respectively, we computed empirical biases of the estimators for areas and subareas as 

\begin{align*}
B(\hat{F}_{\alpha d})=I^{-1}\sum_{i=1}^{I}(\hat{F}_{\alpha d}^{(i)}-{F}_{\alpha d}^{(i)}), \ \ B(\hat{F}_{\alpha dj})=I^{-1}\sum_{i=1}^{I}(\hat{F}_{\alpha dj}^{(i)}-{F}_{\alpha dj}^{(i)}),
\end{align*}
where $\hat{F}_{\alpha d}^{(i)}$ and $\hat{F}_{\alpha dj}^{(i)}$ denote the estimators for the simulated population $i$. Similarly, we computed empirical MSEs of the estimators for areas and subareas as 

\begin{align*}
MSE(\hat{F}_{\alpha d})=I^{-1}\sum_{i=1}^{I}(\hat{F}_{\alpha d}^{(i)}-{F}_{\alpha d}^{(i)})^{2}, \ \ MSE(\hat{F}_{\alpha dj})=I^{-1}\sum_{i=1}^{I}(\hat{F}_{\alpha dj}^{(i)}-{F}_{\alpha dj}^{(i)})^{2}.
\end{align*}

\vspace{7mm}

\noindent\textbf{4.1. \quad ${e}_{djk}$ skew normal}\\

\vspace{2mm}

\begin{figure}[!h]
	\centering
	\begin{subfigure}[b]{0.39\textwidth}
		\centering
		\includegraphics[width=\textwidth]{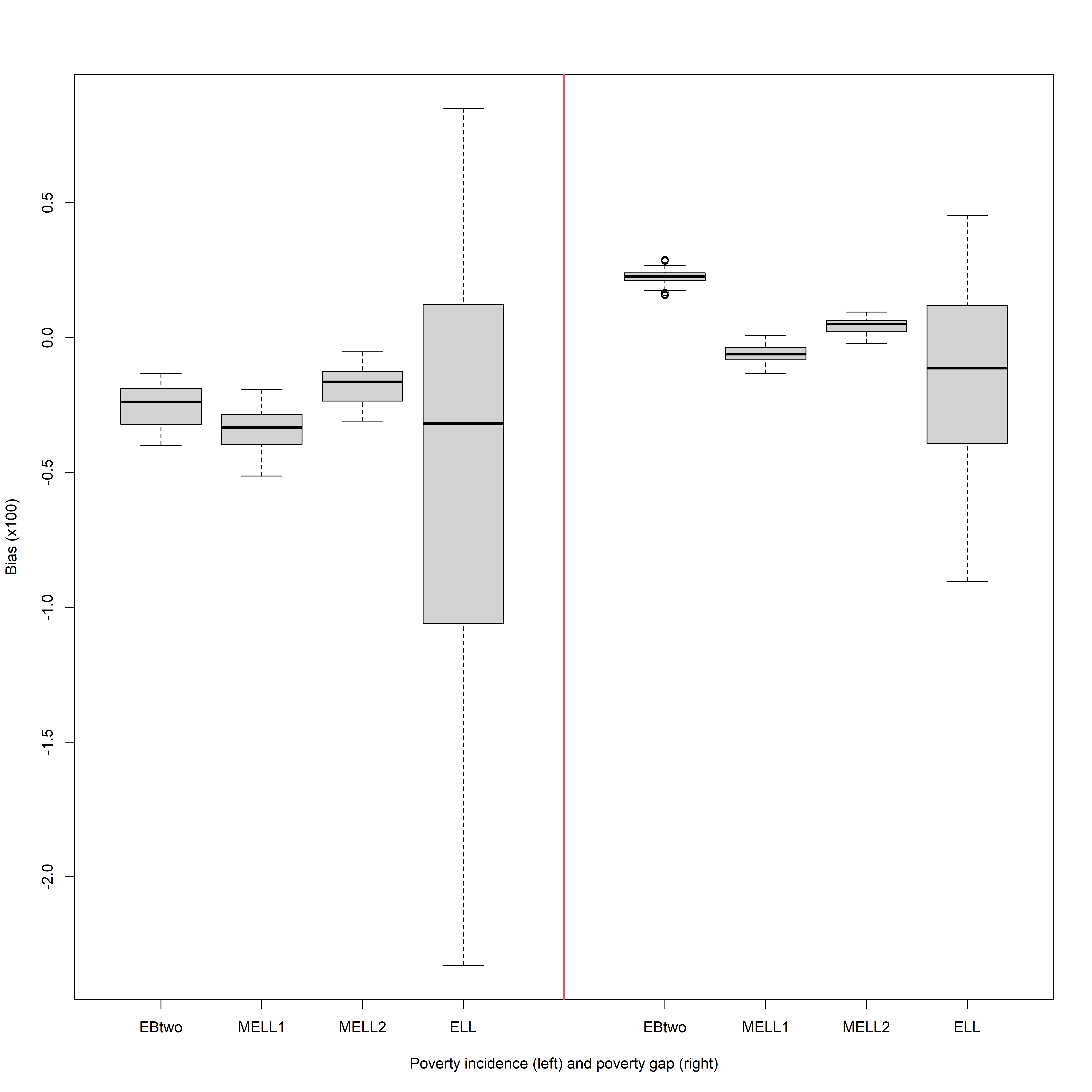}
		\caption{Case 1,  areas}
		\label{fig:gull}
	\end{subfigure}
	\begin{subfigure}[b]{0.39\textwidth}
		\centering
		\includegraphics[width=\textwidth]{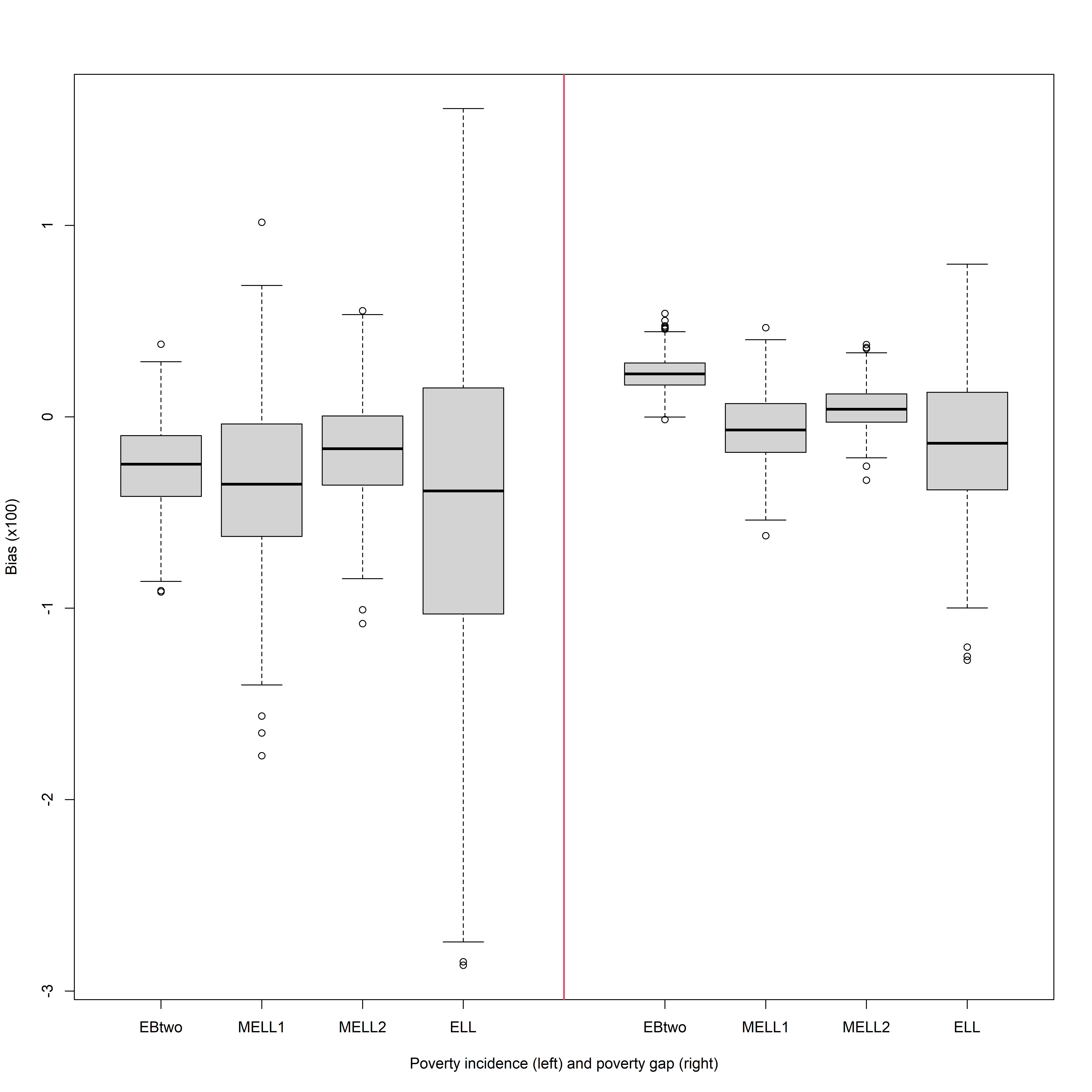}
		\caption{Case 1, subareas}
		\label{fig:tiger}
	\end{subfigure}\\
	\begin{subfigure}[b]{0.39\textwidth}
		\centering
		\includegraphics[width=\textwidth]{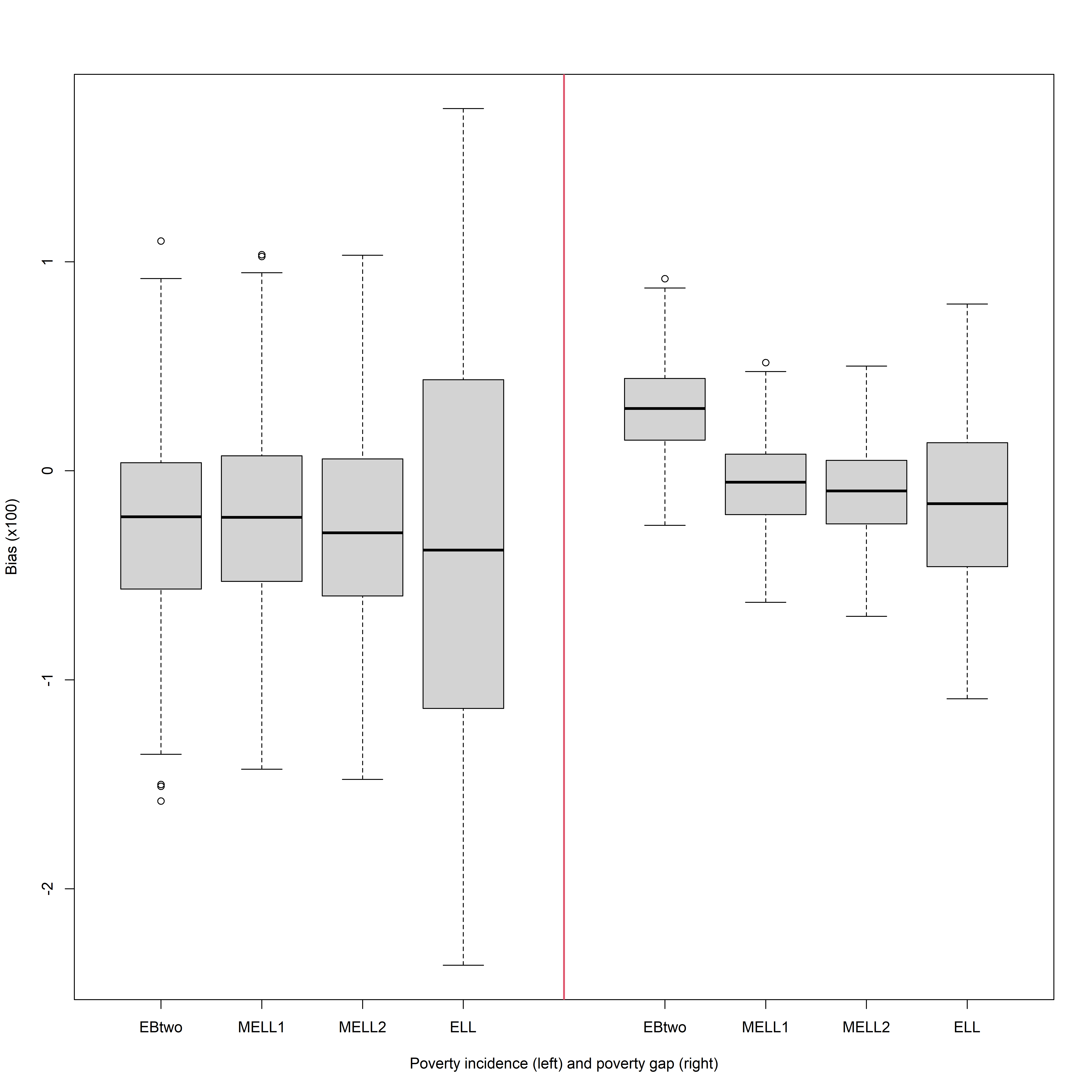}
		\caption{Case 2, non-sampled subareas }
		\label{fig:box1}
	\end{subfigure}
	\caption{Boxplots of biases  $(\times 100)$  over simulated populations of EBtwo, MELL1,  MELL2, and  ELL estimators of  the poverty incidence (left side) and the poverty gap (right side) for the area and subareas in Case 1  are presented in (a)  and (b) and for non-sampled subareas of Case 2 is presented in (c) (${e}_{djk}$ is SN).}
	\label{fig:Bias1}
\end{figure}

Figure \ref{fig:Bias1} presents box plots of bias $(\%)$ for ELL, modified ELL1 (MELL1), modified ELL2 (MELL2), and EBtwo estimators of FGT poverty incidence and poverty gap for areas and subareas under scenario 1 with skew normal errors  ${e}_{djk}$. Here EBtwo denotes empirical best estimator of Marheunda et al. (2017) assuming  $(u_d,v_{dj},e_{djk})$   normal.

Box plots in Figure \ref{fig:Bias1} show that ELL performs significantly worse than the other methods, leading to substantial underestimation in all cases, particularly for poverty incidence. Overall, MELL2 and EBtwo perform better than MELL1 although EBtwo leads to slight overestimation for poverty gap. 

Table \ref{T1} reports results on average MSE for the areas, sampled subareas and non-sampled subareas. Table \ref{T1} shows that ELL leads to very large average MSE in all cases compared to the other methods. For areas, MELL2 and MELL1 are comparable and slightly better than EBtwo in terms of average MSE. For the case where all subareas are sampled (case 1), MELL2 is significantly better than MELL1. This is to be expected because MELL1 does not use subarea specific method unlike MELL2. Also, EBtwo seems to be somewhat  better than MELL2 in terms of average MSE: $8.81$ for EBtwo vs. $11.39$ for MELL2 in the case of poverty gap.

\begin{table}[!h]
\centering
\caption{
Average of MSEs  $(\times 10^4)$. Case 1 when all subareas are sampled and case 2 is when all subareas are not sampled ($e_{djk}$ is SN).}
\label{T1}
\begin{tabular}{lcccccc}
  \hline
  {} & {} & {Poverty} & {} &  Estimation method & {} & {}  \\
    \cline{4-7}
    &  &  indicator &  EBtwo & MELL1 & MELL2 & 
   ELL\\
  \hline
   & Area & inc &  8.70 & 7.88 & 7.28 & 557.44\\
     &  & gap &  1.28 & 1.30 & 1.17 & 95.33\\
     \cline{3-7}
  Case 1 & Subarea  & inc  & 56.09 & 187.69 & 69.35 & 737.22\\
     &  & gap &  8.81 & 33.52 & 11.39 & 127.53\\
       \hline
   & Area & inc &  24.15 & 23.69 & 23.11 & 562.45\\
     &  & gap &  4.22 & 4.27 & 4.17 & 96.46\\
     \cline{3-7}
   & Sampled- & inc &  27.55 & 167.68 & 34.49 & 742.50\\
  Case 2   & subarea  & gap & 3.97 & 29.81 & 5.02 & 128.94\\
     \cline{3-7}
     & Nonsampled-  & inc &  233.85 & 236.05 & 236.59 & 738.95\\
     & subarea & gap & 41.77 & 42.39 & 42.43 & 127.65\\
  \hline
\end{tabular}
\end{table}

Turning to case 2 where not all subareas are sampled, results for sampled subareas are similar those for case 1 where all subareas are sampled. Note that the average MSE is significantly decreased for sampled subareas because the sample size in those subareas is doubled relative to case 1. On the other hand, for areas the average MSE is significantly increased in case 2 compared to case 1 because the number of sampled subareas is reduced by half compared to case 1.     

\begin{figure}[!h]
	\centering
	\begin{subfigure}[b]{0.39\textwidth}
		\centering
		\includegraphics[width=\textwidth]{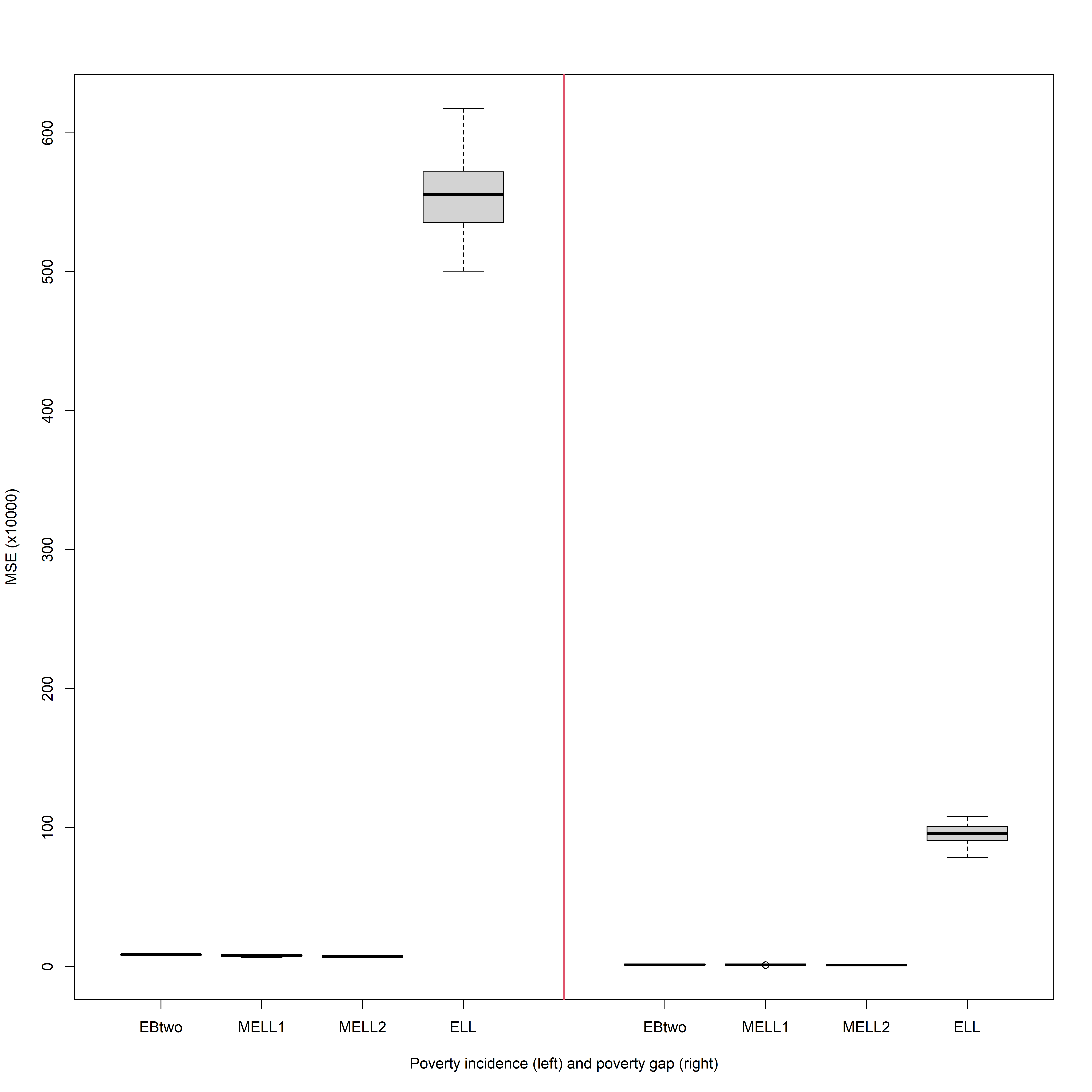}
		\caption{Case 1, MSEs of all areas}
		\label{fig:gull}
	\end{subfigure}
	\begin{subfigure}[b]{0.39\textwidth}
		\centering
		\includegraphics[width=\textwidth]{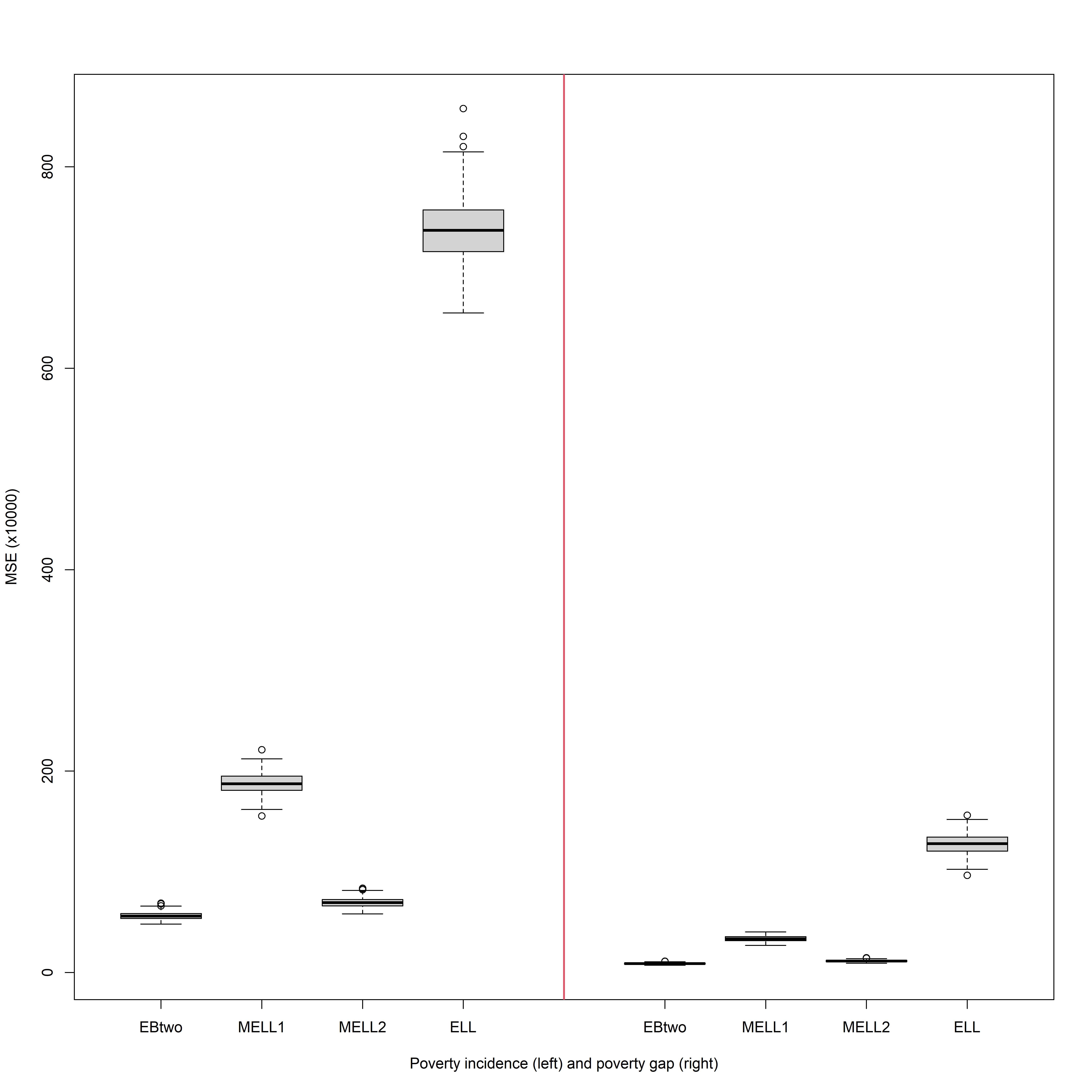}
		\caption{Case 1, MSEs of all subareas}
		\label{fig:tiger}
	\end{subfigure}\\
	\begin{subfigure}[b]{0.39\textwidth}
		\centering
		\includegraphics[width=\textwidth]{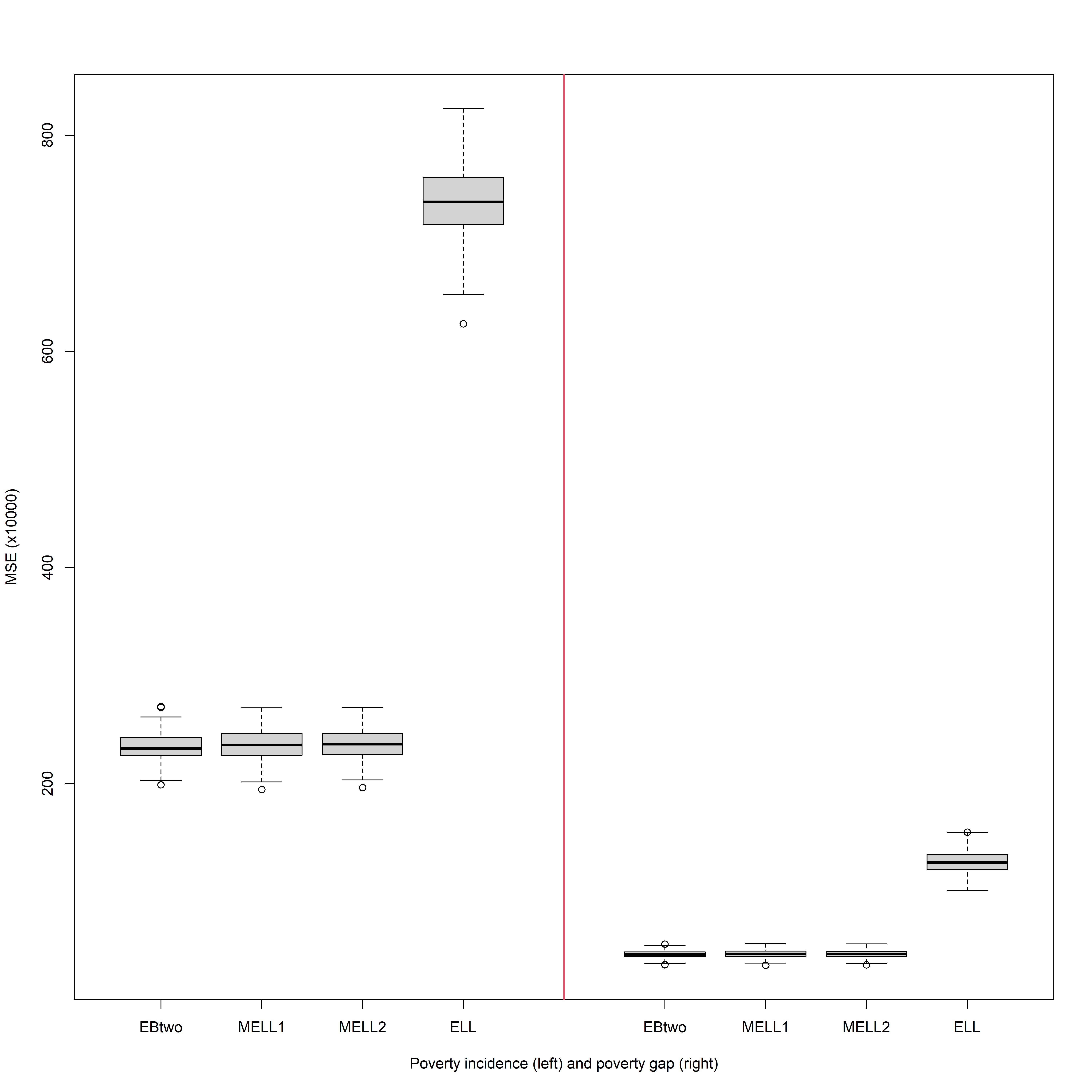}
		\caption{Case 2, MSEs of all non-sampled subareas }
		\label{fig:box1}
	\end{subfigure}
	\caption{Boxplots of MSEs  $(\times 10^4)$  over simulated populations of  EBtwo, MELL1,  MELL2, and  ELL estimators of  the poverty incidence (left side) and the poverty gap (right side) for the area and subareas in Case 1  are presented in (a)  and (b) and for non-sampled subareas of Case 2 is presented in (c) (${e}_{djk}$ is SN).}
	\label{fig:MSE1}
\end{figure}

For nonsampled subareas (case 2), MELL1, MELL2 and EBtwo are comparable in terms of average MSE. This is expected because for non-sampled subareas MELL1 and MELL2 are similar. Note that the average MSE is significantly increased for nonsampled subareas compared to corresponding values for sampled subareas. Figure  \ref{fig:MSE1} presents box plots of MSE for areas, sampled subareas and non-sampled subareas. Conclusions from those plots are like those arrived from the values of average MSE.

\vspace{5mm}

\noindent\textbf{4.2. \quad $({v}_{dj},{e}_{djk})$ skew normal }\\

\vspace{2mm}

\begin{figure}[!h]
	\centering
	\begin{subfigure}[b]{0.39\textwidth}
		\centering
		\includegraphics[width=\textwidth]{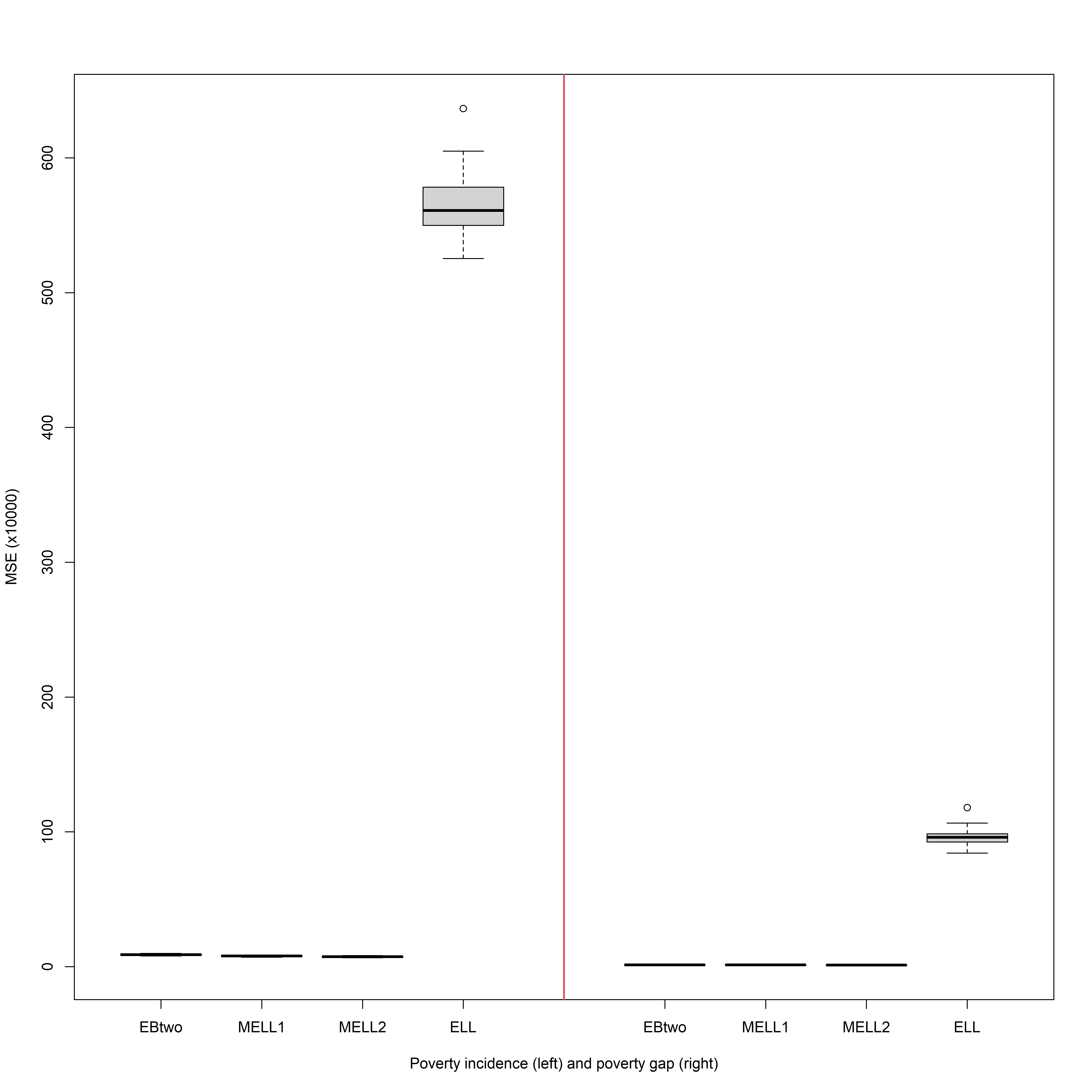}
		\caption{Case 1, MSEs of all areas}
		\label{fig:gull}
	\end{subfigure}
	\begin{subfigure}[b]{0.39\textwidth}
		\centering
		\includegraphics[width=\textwidth]{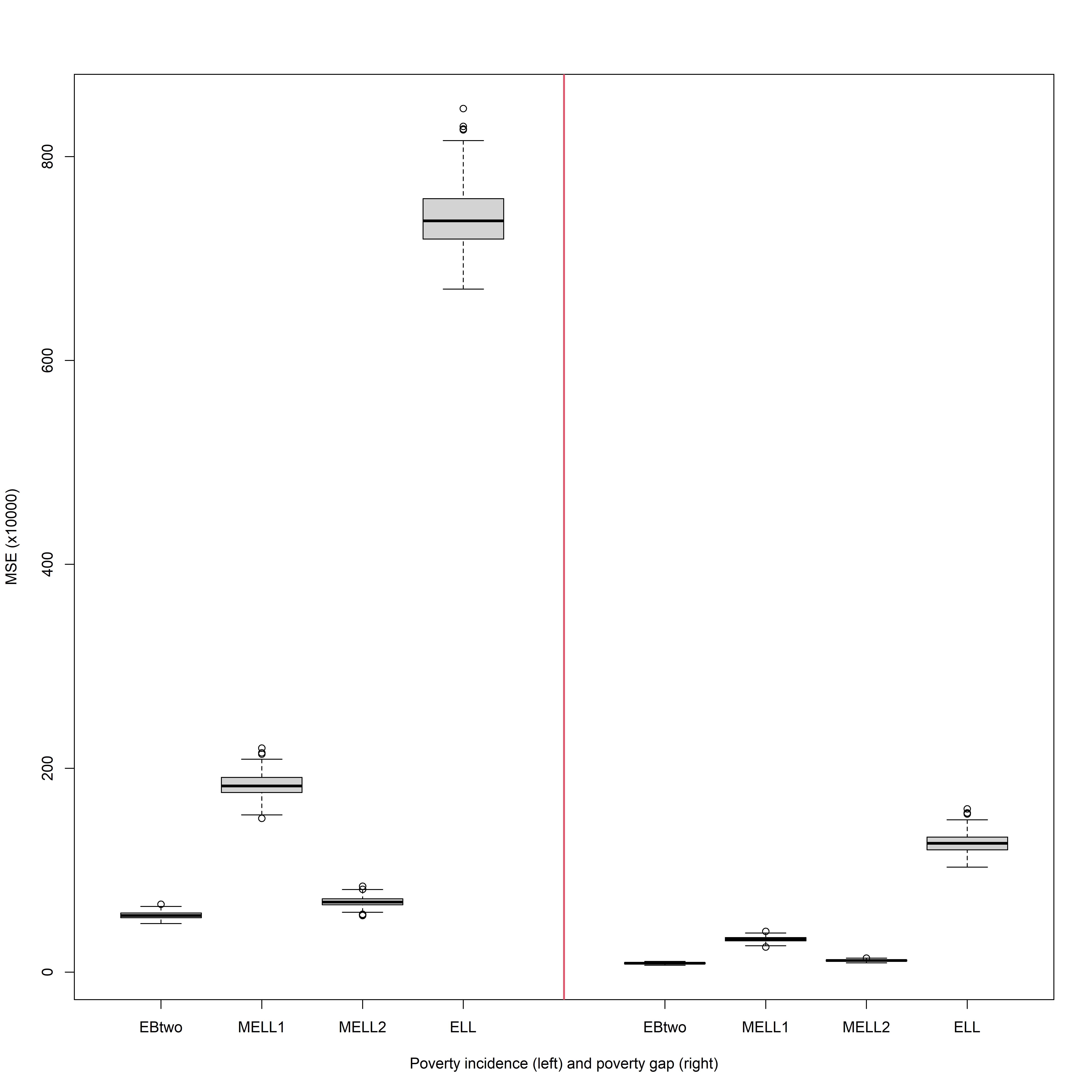}
		\caption{Case 1, MSEs of all subareas}
		\label{fig:tiger}
	\end{subfigure}\\
	\begin{subfigure}[b]{0.39\textwidth}
		\centering
		\includegraphics[width=\textwidth]{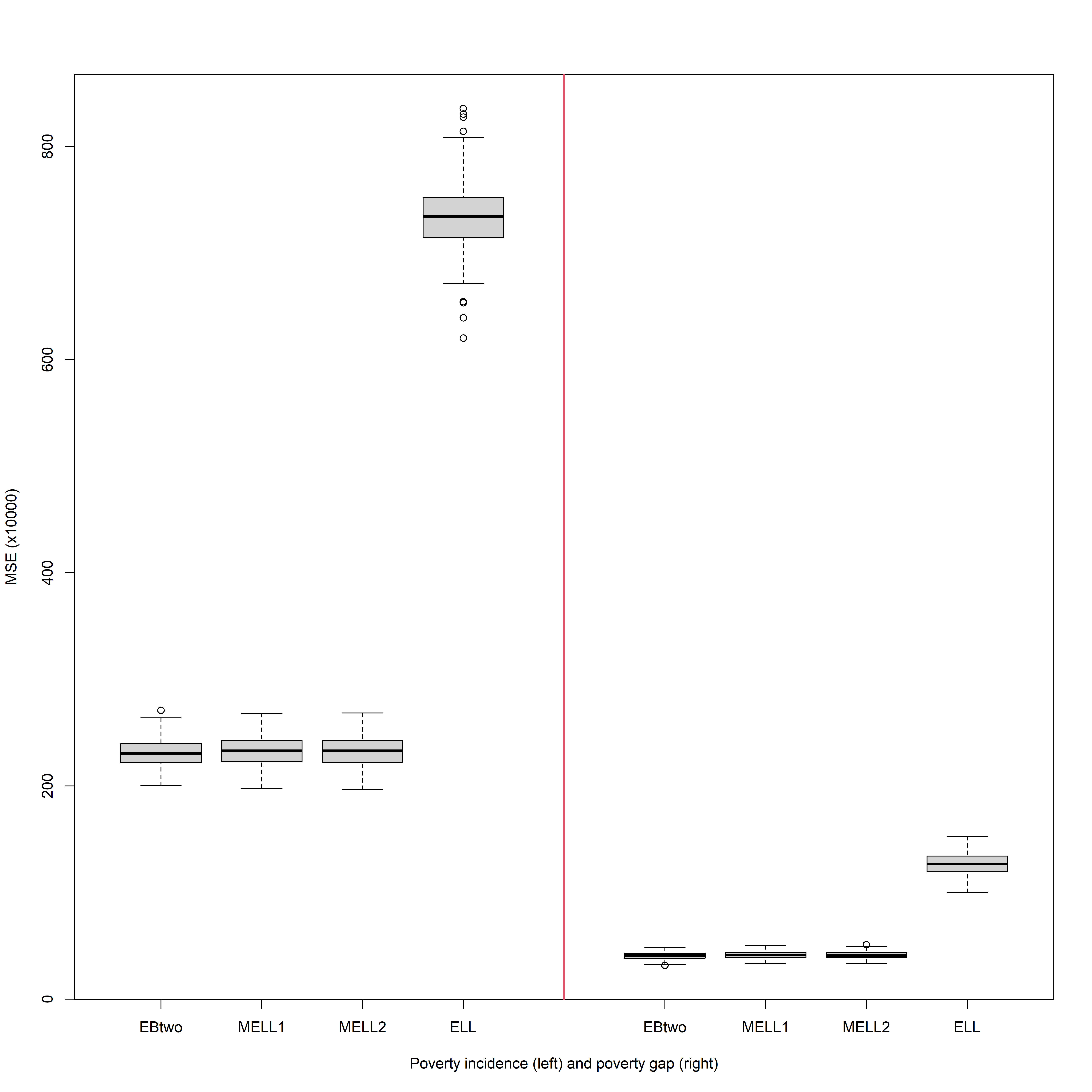}
		\caption{Case 2, MSEs of all non-sampled subareas }
		\label{fig:gull}
	\end{subfigure}
	\caption{Boxplots of MSEs  $(\times 10^4)$  over simulated populations of  ELL, MELL1,  MELL2, and EBtwo estimators of  the poverty incidence (left side) and the poverty gap (right side)  for the area and subareas in Case 1  are presented in (a)  and (b) and for non-sampled subareas of Case 2 is presented in (c) (${v}_{dj}$ and ${e}_{djk}$ is SN).}
	\label{fig:MSE2V}
\end{figure}

We also considered the case where both ${v}_{dj}$  and ${e}_{djk}$ are skew normal and ${u}_{d}$ normal. Average MSE and Box plots of MSE for areas, sampled subareas and nonsampled subareas, reported in Table  \ref{T2} and Figure  \ref{fig:MSE2V} respectively, are very similar to those reported for the case where only ${e}_{djk}$  is SN. Therefore, our conclusions for the two cases are similar.

\begin{table}[!h]
\centering
\caption{
Average of MSEs  $(\times 10^4)$. Case 1 when all subareas are sampled and case 2 is when all subareas are not sampled ($v_{dj}$ and $e_{djk}$ are SN).}
\label{T2}
\begin{tabular}{lcccccc}
  \hline
  {} & {} & {Poverty} & {} & Estimation method & {} & {}  \\
    \cline{4-7}
    &  &  indicator &EBtwo & MELL1 & MELL2 & 
   ELL\\
  \hline
   & Area & inc  & 8.89 & 7.92 & 7.36 & 564.12\\
     &  & gap &  1.32 & 1.31 & 1.19 & 95.71\\
     \cline{3-7}
  Case 1 & Subarea  & inc &  55.79 & 182.97 & 69.11 & 739.27\\
     &  & gap &  8.74 & 32.29 & 11.43 & 126.68\\
      \hline
   & Area & inc &  23.76 & 23.48 & 22.76 & 561.67\\
     &  & gap &  4.13 & 4.23 & 4.09 & 96.70\\
     \cline{3-7}
   & Sampled- & inc &  27.28 & 164.16 & 34.18 & 739.02\\
 Case 2    & subarea  & gap &  3.91 & 28.82 & 5.00 & 128.50\\
     \cline{3-7}
     & Nonsampled-  & inc &  230.19 & 232.75 & 232.49 & 734.29\\
     & subarea & gap &  40.69 & 41.35 & 41.27 & 126.62\\
  \hline
\end{tabular}
\end{table}

\vspace{5mm}

\noindent\textbf{4.3. \quad $({u}_{d}$, ${v}_{dj}$, ${e}_{djk})$ Normal }\\

\vspace{2mm}

We also considered the case where   ${u}_{d}$, ${v}_{dj}$ and ${e}_{djk}$   are normally distributed. Marheunda et al. (2017) studied this case in the context of EBtwo estimators. Again our results on MSE for areas, sampled subareas and non-sampled subareas indicate similarity with the results in sections 4.1 and 4.2 corresponding to  ${e}_{djk}$ skew normal and $({v}_{dj},{e}_{djk})$ skew normal. We report results only on average MSE in Table \ref{T3}. We note that EBtwo leads to substantial reduction in average MSE over MELL2 for subareas in case 1 where all subareas are sampled: $49.77$ vs. $64.30$ for incidence and $8.84$ vs. $11.38$ for gap. This is to be expected because EBtwo is optimal under normality.

\begin{table}[!h]
\centering
\caption{
Average of MSEs  $(\times 10^4)$. Case 1 when all subareas are sampled and case 2 is when all subareas are not sampled (${u}_{d}$, ${v}_{dj}$ and ${e}_{djk}$ all N).}
\label{T3}
\begin{tabular}{lccccccc}
  \hline
  {} & {} & {Poverty} & {} & Estimation method & {} & {}  \\
    \cline{4-7}
    &  &  indicator &  EBtwo & MELL1 & MELL2 & 
   ELL\\
  \hline
   & Area & inc &	5.87 &	6.82 &	6.42 &	508.88\\
     &  & gap &  1.07 & 1.23 & 1.14 & 93.07\\
     \cline{3-7}
  Case 1 &  Subarea & inc & 	49.77 &	168.60 &	64.30 &	670.71\\
     & & gap &  8.84 & 32.14 & 11.38 & 123.98\\
      \hline
   & Area & inc &  19.92 & 21.73 & 21.32 & 514.96\\
     &  & gap & 3.81 & 4.11 & 4.03 & 94.63\\
     \cline{3-7}
  & Sampled- & inc & 24.62 & 151.31 & 31.31 & 677.23\\
  Case 2    & subarea  & gap &  4.11 & 29.04 & 4.83 & 126.31\\
     \cline{3-7}
     & Nonsampled-  & inc &  209.02 & 214.87 & 215.08 & 675.51\\
     & subarea & gap &  40.22 & 41.28 & 41.32 & 125.06\\
  \hline
\end{tabular}
\end{table}

\vspace{5mm}

\noindent\textbf{4.4. \quad Comparison with one-fold model }\\

\vspace{2mm}

In this section, we study the effect of ignoring the area effect and use a one-fold model containing only subarea random effects. In particular, we are interested in the performance of MSE for non-sampled subareas when the true model is the two-fold model. Denoting the subareas by a single index  $t$, the one-fold model may be written as ${y}_{ tk}=\mathbf{x}_{tk}^{'}\boldsymbol{\beta}+{u}_{t}+{e}_{tk}$, like the model studied by Diallo and Rao (2017). We can then use their results to get ELL estimators for non-sampled subareas. They have also studied modified ELL under the one-fold model, but for non-sample subareas it is essentially the same as ELL. For sampled subareas, modified ELL was shown to be more efficient than ELL under the one-fold model. 

\begin{figure}
	\centering
		\includegraphics[width=0.39\textwidth]{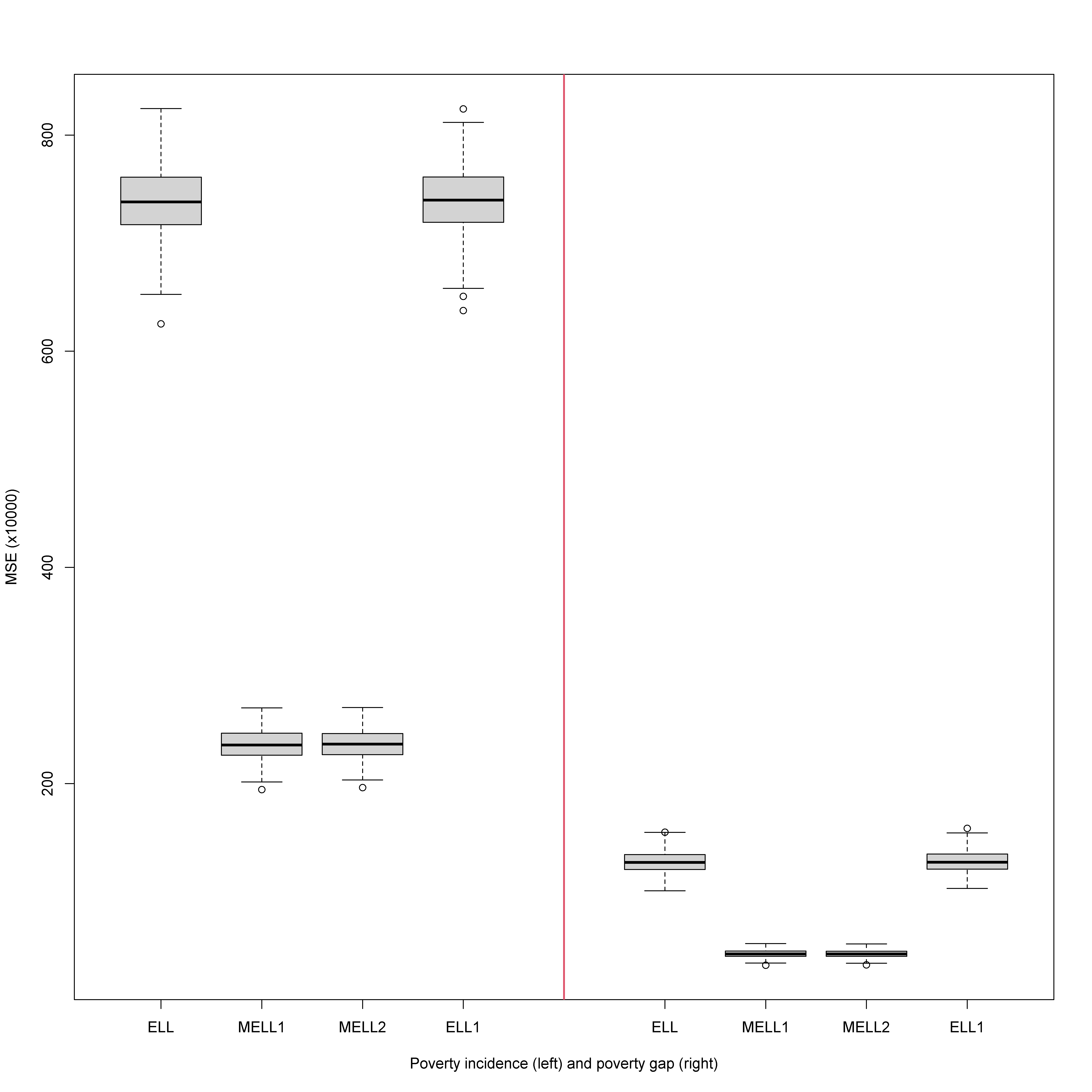}
	\caption{Boxplots of MSEs  $(\times 10^4)$  over simulated populations of  two fold and one fold  estimators of  the poverty incidence (left side) and the poverty gap (right side) for each non-sampled subarea of Case 2 ( ${e}_{djk}$ is SN).}.
	\label{fig:MSENSTO}
\end{figure}

We used the case 2 set up of our simulation study and fitted the one-fold model to the sample observations and obtained ELL estimators for nonsampled subareas for each simulated sample. Resulting box plots of MSE of ELL based on one-fold model, denoted ELL1, and ELL, MELL1 and MELL 2 based on two-fold model for the nonsampled subareas are reported in Figure \ref{fig:MSENSTO}. Average MSE values are reported in Table \ref{T4}.

It is clear from the box plots and average MSE values that MELL1 and MELL2 behave similarly in terms of MSE  and lead to large reduction in MSE relative to ELL1 and ELL. We also note that ELL based on the two-fold model and ELL1 based on the one-fold model give similar results in terms of MSE.

\begin{table}[!h]
\centering
\caption{
Average of MSEs  $(\times 10^4)$ of all non-sampled  subareas ($e_{djk}$ is SN)}
\label{T4}
\begin{tabular}{lcccccc}
  \hline
  {} & {} & {Poverty} & {} & Estimation method & {} & {}  \\
    \cline{4-7}
    &  &  indicator &ELL & MELL1 & MELL2 & 
   ELL1\\
  \hline     
& Nonsampled-  & inc &  738.95 & 236.05 & 236.59 & 740.31\\
     & subarea & gap & 127.65 & 42.39 & 42.43 & 127.94\\
  \hline
\end{tabular}
\end{table}

\vspace{5mm}

\begin{center}
\section{MSE ESTIMATION}
\end{center}

In the ELL method for the one-fold model, the variability of the simulated census measures is taken as the estimator of MSE of the ELL estimator. Similarly, under the two-fold model the corresponding MSE estimators of ELL for areas and subareas are given by 

\begin{align}
\label{MSE51}
MSE(\hat{F}_{\alpha d}^{ELL})=B^{-1}\sum_{i=1}^{B}({F}_{\alpha d}^{*(b)}-\hat{F}_{\alpha d}^{ELL})^{2} \ \ \end{align}

and

\begin{align}
\label{MSE52}
MSE(\hat{F}_{\alpha dj}^{ELL})=B^{-1}\sum_{i=1}^{B}({F}_{\alpha dj}^{*(b)}-\hat{F}_{\alpha dj}^{ELL})^{2} .
\end{align}
MSE estimators similar to \eqref{MSE51} and \eqref{MSE52} are applicable to MELL1 and MELL2, using simulated census measures. The proposed MSE estimators are simple, but they can lead to significant underestimation of the true MSE because the model parameters and the random effects in the model are not re-estimated in each replicate   from the replicated sample data  $({y}_{ djk}^{*(b)},\mathbf{x}_{djk})$. 

Marheuda et al. (2017) proposed a proper parametric bootstrap MSE estimator for EBtwo estimators, based on re-estimating model parameters and random effects in the two-fold model under normality. A similar procedure may be developed for ELL and MELL using a distribution free bootstrap, like the ELL method. 

\vspace{5mm}

\begin{center}
\section{CONCLUDING REMARKS}
\end{center}

We considered the estimation of FGT poverty measures under a two-fold nested error model. We developed extensions of the ELL method and the modified ELL method of Diallo and Rao (2017) to two-fold models. The methods are free of parametric distributional assumptions on the random effects in the two-fold model.  Our simulation results indicate that the proposed modified ELL methods lead to large efficiency gains over the ELL for both areas and subareas. Further, MELL2 leads to significant reduction in MSE over MELL1 for sampled subareas, and it is comparable to the EBtwo method of Marheuda et al. (2017) under normality assumption. An advantage of MELL2 is that it is applicable to more complex parameters, not necessarily additive in the individual values like FGT measures, unlike EBtwo. 
Bootstrap MSE estimation for MELL methods, along the lines of Marheuda et al. (2017) but without normality assumption, needs a detailed investigation.


\vspace{5mm}


\begin{thebibliography}{}
	

\bibitem{BHF}  Battese, G.E., Harter, R.M. and Fuller, W.A. (1988). An error-components 	model for prediction of county crop areas using survey and satellite data. \emph{Journal of American Statistical Association} \textbf{83} 28 - 36.

\bibitem{DR}   Diallo, M.S. and  Rao, J. N. K. (2018). Small area estimation of complex parameters
under unit-level models with skew-normal errors. \emph{Scandinavian Journal of Statistics} \textbf{45} 1092 - 1116.

\bibitem{ELL}   Elbers, C., Lanjouw, J.O. and Lanjouw, P. (2001).  \emph{Welfare in villages and towns: micro-level
estimation of poverty and inequality}. Unpublished manuscript, The World Bank.

\bibitem{FGT} Foster, J., Greer, J., \& Thorbecke, E. (1984). A class of decomposable poverty measures. \emph{Econometrica} \textbf{52} 761 – 766.


\bibitem{HDCI} Hossain, J., Das, S, Chandra, H. and Islam, M. A. (2020). Disaggregate level estimates and spatial mapping of food insecurity in Bangladesh by linking survey and census data. \emph{Plos One} \textbf{15} 1 – 16.

\bibitem{MMMR} Marhuenda, Y., Molina, I., Morales, D., \& Rao, J. N. K. (2017). Poverty mapping in small areas under a twofold nested error regression model. \emph{Journal of the Royal Statistical Society: Series A} \textbf{180} 1111 – 1136.
\bibitem{MR}   Molina, I. and  Rao, J. N. K. (2010). Small area estimation of poverty indicators. \emph{Cananadian Journal of 
Statistics} \textbf{38} 369 - 385.

\bibitem{RM} Rao, J. N. K., \& Molina, I. (2015). \emph{Small-area estimation} (2nd ed.). Hoboken, NJ: Wiley.



\end{thebibliography}
\end{document}